\documentclass[aps,prb,amsmath,amssymb,twocolumn,showpacs]{revtex4}
\usepackage{bm}
\usepackage{graphicx}
\usepackage{epstopdf}
\usepackage{color}  
\newcommand{\nix}[1]{}
\begin{document}
\title{Weak localization of holes in high-mobility heterostructures}
\author{F. V. Porubaev}
\author{L. E. Golub}
\email{golub@coherent.ioffe.ru}
\affiliation{Ioffe Physical-Technical Institute of the Russian Academy of Sciences, 194021 St.~Petersburg, Russia}%

\begin{abstract}

Theory of weak localization is developed for two-dimensional holes in semiconductor heterostructures. Ballistic regime of weak localization where the backscattering occurs from few impurities is studied with account for anisotropic momentum scattering of holes. The transition from weak localization to anti-localization is demonstrated for long dephasing times. For stronger dephasing the conductivity correction is negative at all hole densities due to non-monotonous dependence of the spin relaxation time on the hole wavevector. The anomalous temperature dependent correction to the conductivity is calculated.
We show that the temperature dependence of the conductivity is non-monotonous at moderate hole densities.
\end{abstract}

\maketitle

Weak localization (WL) is an enhancement of backscattering caused by
interference of waves propagating by the same path in the opposite directions. WL of electrons results in corrections to conductivity anomalously dependent on temperature and magnetic field. Weak localization is studied in a variety of disordered metals and semiconductors including recently discovered graphene and topological insulators. 
In these systems, the carrier momentum is coupled strongly with spin or pseudospin which results in non-parabolic energy spectrum.
This leads to suppression of the backscattering called weak antilocalization (WAL). WAL is observed in graphene, for review see Ref.~\onlinecite{Sarma_review}, and in topological insulators,~\cite{topins_exp1,topins_exp2,topins_exp3} the corresponding theory is developed in Refs.~\onlinecite{topins_theor1,topins_theor2}.

Heterostructures with hole type of conductivity also belong to the class of systems with nonparabolic energy spectrum and coupled spin and orbital degrees of freedom. Since the two-dimensional valence band states are formed as a result of competition of size quantization and strong spin-orbit interaction present in the bulk semiconductor, the spectrum nonparabolicity depends on the two-dimensional hole density $p$ and the quantum well width $a$.
At small density and in narrow structures 
the holes behave as ordinary electrons, while at 
$pa^2 \geq 1$ 
the nonparabolicity effects are important and spin is strongly coupled with the momentum.
This leads to transition from WL at low density to WAL at higher densities in 2D hole systems caused by increase of spin-orbit strength.~\cite{SOI_MIT}

Experimental studies showed the WL-WAL transition with increase of 2D hole density
in both temperature and magnetic-field dependences of the conductivity.~\cite{MIT_holes_1,Minkov_strained,Minkov_asymmetric}  
The WL theory for hole systems~\cite{JETP98,SSC98,FTP98} was developed for the so-called ``diffusion'' regime where the backscattering occurs after many collisions with impurities. The diffusion theory describes well the magnetoconductivity in low-mobility hole structures,~\cite{Danemark} but it is inapplicable to high-mobility systems studied nowadays because it describes magnetoconductivity in very low fields. 

The temperature dependence of the conductivity is also different in high-mobility systems. WL is present 
due to these closed paths where the interference is not broken by dephasing processes. This means that the phase breaking time $\tau_\phi$ should be much longer than the momentum scattering time $\tau$. However, the diffusion theory of WL assumes a stronger condition $\ln{(\tau_\phi/\tau)} \gg 1$ which is not realized in high-mobility structures with long scattering times. The backscattering in these systems occurs after carrier propagation along  ``ballistic'' paths with few impurities which are ignored by the diffusion theory.
The aim of this work is to develop the WL theory of two-dimensional holes valid in both the diffusion and ballistic regimes and calculate the temperature dependence of the conductivity correction.

\begin{figure}[tb]
\includegraphics[width=\linewidth]{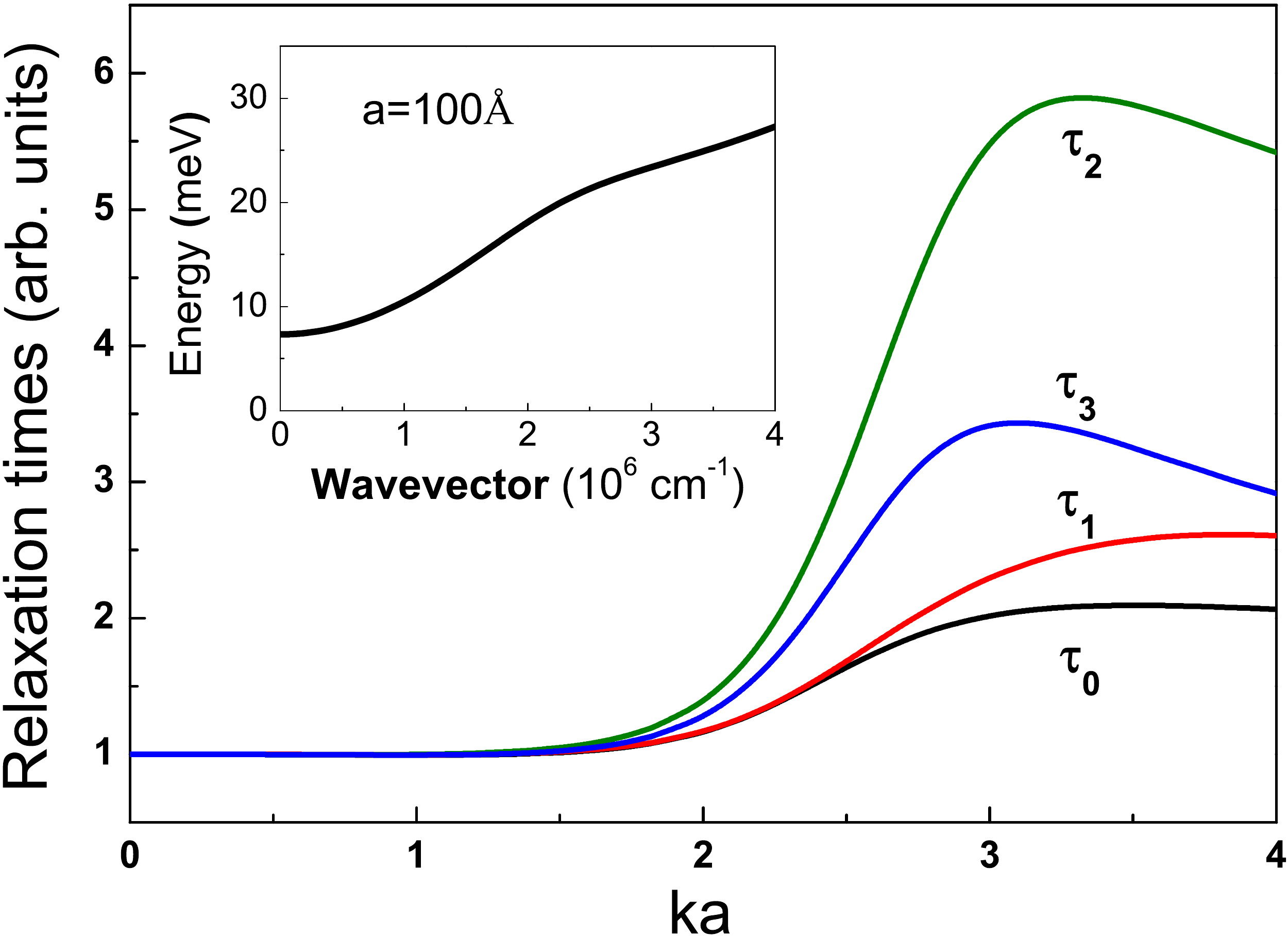} 
\caption{Elastic relaxation times of  holes. Inset: energy dispersion of the ground hole subband in a 100~\AA \,  wide quantum well.}
\label{fig_disp_times}
\end{figure}

The hole states in quantum-well structures grown from A$_3$B$_5$ materials are described by the Luttinger Hamiltonian where the wavevector component along the growth direction  is an operator. We apply the isotropic approximation and assume the quantum well to be rectangular and infinitely deep. We have found the hole energy dispersion in the subbands of size quantization in this model by a standard way.~\cite{EL_GE_book} 
The dependence of the hole energy in the ground subband on the two-dimensional wavevector, $E(k)$, is  non-parabolic at $ka \geq 2$, see inset to Fig.~\ref{fig_disp_times}.

The size-quantized hole levels in a symmetrical quantum well are double degenerate at each value of  $\bm k$. We choose these two states as symmetric ($s$) and asymmetric ($a$) relative to a mirror reflection in the symmetry plane of the quantum well.~\cite{MPP_91} The corresponding wave functions are $\Psi_{\alpha \bm k} = F_{\alpha \bm k}(z) \exp{({\rm i}\bm k \cdot \bm \rho)}$, where $z$ is the direction of quantization, $\bm \rho$ is the in-plane coordinate, $\alpha=s,a$, and $F_{\alpha \bm k}(z)$ are 
linear superpositions of the four Bloch functions $u_m$ of the total angular momentum 3/2 and projections on the growth axis $m=\pm 3/2, \pm 1/2$:
$u_{\pm 3/2}= \mp (X\pm {\rm i}Y) s_\pm/\sqrt{2}$, 
$u_{\pm 1/2}= [\mp (X\pm {\rm i}Y) s_\mp/\sqrt{2} + \sqrt{2} Z s_\pm]/\sqrt{3}$.
Here $X,Y,Z$ are the Bloch functions of the top of the valence band, and $s_\pm$ are spin functions with  projections $\pm1/2$ on the $z$ axis. 
In the basis $[u_{3/2}, u_{1/2}, u_{-1/2}, u_{-3/2}]$, $F_{\alpha \bm k}$  have the following form~\cite{MPP_91,JETP98}
\begin{eqnarray}
\label{s}
&&	F_{s \bm k}(z) = \\
&&	\left[-v_0 C(z), {\rm i}v_1 S(z) {\rm e}^{{\rm i}\varphi_{\bm k}}, -v_2 C(z){\rm e}^{2{\rm i}\varphi_{\bm k}}, {\rm i}v_3 S(z) {\rm e}^{3{\rm i}\varphi_{\bm k}}\right],
	\nonumber \\
\label{a}
&&	F_{a \bm k}(z) = \\
&&	\left[{\rm i}v_3 S(z) {\rm e}^{-3{\rm i}\varphi_{\bm k}}, v_2 C(z){\rm e}^{-2{\rm i}\varphi_{\bm k}}, {\rm i}v_1 S(z) {\rm e}^{-{\rm i}\varphi_{\bm k}}, v_0 C(z) \right],
	\nonumber
\end{eqnarray}
where $\varphi_{\bm k}$ is the polar angle of the wavevector $\bm k$. The dependence on $k$ of the real coefficients $v_{0\ldots 3}$ as well as symmetric ($C$) and antisymmetric ($S$) functions of the coordinate $z$ is determined by the energy dispersion $E(k)$.

We consider elastic scattering of two-dimensional holes by the short-range potential $V(\bm r)=V_0\sum_i \delta(\bm r - \bm r_i)$, where $\bm r_i = (\bm \rho_i,z_i)$ are coordinates of the impurities. 
The spin-orbit coupling leads to anisotropy of scattering even by such isotropic impurities. Indeed, the hole wavefunctions Eqs.~\eqref{s},~\eqref{a} depend on a direction of
the wavevector $\bm k$. 
As a result, the  matrix elements of scattering from the state $|\alpha \bm k\rangle$ to the state $|\beta \bm k'\rangle$ by the potential $V(\bm r)$ 
\[
V_{\beta\alpha}(\varphi_{\bm k'},\varphi_{\bm k})
= V_0 \sum_i F_{\beta \bm k'}^\dag (z_i) F_{\alpha \bm k}(z_i) {\rm e}^{{\rm i}(\bm k - \bm k')\cdot \bm \rho_i}
\]
depend on the angular coordinates of the initial and final wavevectors $\varphi_{\bm k}$ and $\varphi_{\bm k'}$. 
%at the Fermi circle. 
Therefore the  elastic scattering
probability contains the 1st, 2nd and 3rd Fourier harmonics of the scattering angle $\theta=\varphi_{\bm k'}-\varphi_{\bm k}$ in addition to the isotropic term. Hence the scattering
 times describing decay of the $n$th Fourier harmonics of the hole distribution function
\[
{1\over \tau_n} = {2\pi \over \hbar} g \int\limits_0^{2\pi} {d\theta \over 2\pi} \left< |V_{ss}(\theta)|^2 + |V_{as}(\theta)|^2 \right> (1-\cos{n\theta})
\]
are different for $n=1,2,3$. All higher harmonics decay with the time $\tau_0$ given by
\[
{1\over \tau_0} = {2\pi \over \hbar} g \int\limits_0^{2\pi} {d\theta \over 2\pi} \left< |V_{ss}(\theta)|^2 + |V_{as}(\theta)|^2 \right>.
\]
Here angular brackets denote averaging over the impurity positions, 
and $g=k/(2\pi \hbar v_k)$ is the density of states where $v_k=\hbar^{-1}dE/dk$ is the velocity of holes with the wavevector $k$.
The dependences $\tau_n(k)$ for $n=0,1,2,3$ are plotted in Fig.~\ref{fig_disp_times} for homogeneous impurity distribution over the $z$ coordinate (averaging is restricted to integration over $z_i$).
Figure~\ref{fig_disp_times}
demonstrates that the $p$-type heterostructures differ qualitatively from topological insulators and graphene where scattering is also anisotropic, but the ratios $\tau_1/\tau_0=2$, $\tau_n/\tau_0=1$  (for $n>1$) are fixed, and all relaxation times are independent of $k$. In contrast, the spin-orbit coupling is absent for holes at $k\to 0$, where the spectrum is parabolic and scattering is isotropic, while at $ka \gtrsim 2$ both conditions are violated.  This implies the spin-orbit coupling strength in the hole systems is controlled by the density $p$ which sets the Fermi wavevector $k_{\rm F}=\sqrt{2\pi p}$.

This anisotropy of scattering complicates the calculation of the WL conductivity correction. It is determined by the Cooperon $C^{\alpha\beta}_{\gamma\delta}(\varphi_{\bm k},\varphi_{\bm k'},\bm q)$ which depends on the four indexes enumerating the hole states ($\alpha,\beta,\gamma,\delta=s,a$), on the directions of the momenta at the Fermi circle $\varphi_{\bm k},\varphi_{\bm k'}$, and on the two-dimensional vector $\bm q$, $q \ll k_{\rm F}$. The Cooperon satisfies the equation~\cite{JETP98,Romanov_Averkiev}
\begin{eqnarray}
\label{C}
&&	C^{\alpha\beta}_{\gamma\delta}(\varphi_{\bm k},\varphi_{\bm k'},\bm q) = 
	W^{\alpha\beta}_{\gamma\delta}(\varphi_{\bm k},\varphi_{\bm k'}) 
	\\
&&	+ 
	\int\limits_0^{2\pi} {d \varphi_1\over 2\pi} P(\varphi_1,\bm q) \, W^{\alpha\mu}_{\gamma\nu}(\varphi_{\bm k},\varphi_1) C^{\mu\beta}_{\nu\delta}(\varphi_1,\varphi_{\bm k'},\bm q),
\nonumber
\end{eqnarray}
where
\[
P(\varphi_1,\bm q) =  {2\pi g_{\rm F}\tau_0 / \hbar \over 1 -{\rm i}qv_{\rm F}\tau_0 \cos{(\varphi_1-\varphi_{\bm q})}},
\]
%is the Fourier transformation of the probability for ballistic propagation.
$g_{\rm F}$ and $v_{\rm F}$ are the density of states and the velocity at $k=k_{\rm F}$, and the correlators are defined as
\begin{equation}
W^{\alpha\beta}_{\gamma\delta}(\varphi_{\bm k},\varphi_{\bm k'}) = \left< V_{\alpha\beta}(\varphi_{\bm k}+\pi, \varphi_{\bm k'}+\pi) V_{\gamma\delta}(\varphi_{\bm k},\varphi_{\bm k'}) \right>.
\end{equation}
For scattering in a symmetrical quantum well the correlators have the properties
following from Eqs.~\eqref{s},\eqref{a}:
\begin{eqnarray}
\label{W}
	W^{ss}_{ss} = (W^{aa}_{aa})^*, 
	\qquad
	W^{ss}_{aa} = W^{aa}_{ss}, 
\\	
W^{sa}_{sa} = (W^{as}_{as})^*, 
	\qquad
	W^{sa}_{as} = W^{as}_{sa},
\nonumber
\end{eqnarray}
and all other correlators are equal to zero. Note that the relaxation times are determined by 
$\left< |V_{ss}|^2 + |V_{as}|^2 \right> = W^{ss}_{aa}-W^{sa}_{as}$. The correlators contain Fourier harmonics $\exp{(\pm{\rm i}m\varphi_{\bm k}\pm{\rm i}n\varphi_{\bm k'})}$ with $m,n=0\ldots 6$, and $W^{ss}_{ss},W^{ss}_{aa},W^{sa}_{as}$ depend on the difference $\varphi_{\bm k}-\varphi_{\bm k'}$ while $W^{sa}_{sa}(\varphi_{\bm k},\varphi_{\bm k'})$ does not.

The correlator properties Eqs.~\eqref{W} allowed us to decouple partly Eqs.~\eqref{C} and derive independent equations for the Cooperons $C_+$ and $C_-$ as well as two coupled equations for the Cooperons $C_1$ and $C_2$
defined as
\begin{eqnarray}
&& C_+ = C^{ss}_{aa}-C^{as}_{sa} = -(C^{sa}_{as}-C^{aa}_{ss}),\\
&& C_-=C^{ss}_{aa}+C^{as}_{sa} = C^{sa}_{as}+C^{aa}_{ss},
\nonumber \\
&& C_1 = C^{ss}_{ss} = (C^{aa}_{aa})^*,
\qquad
C_2 = C^{as}_{as} = (C^{sa}_{sa})^*.
\nonumber 
\end{eqnarray}
The equations are as follows:
\begin{eqnarray}
\label{C_pm}
&&	C_\pm(\varphi_{\bm k},\varphi_{\bm k'},\bm q) = W^{ss}_{aa}(\theta) \mp W^{sa}_{as}(\theta) \\
	&&+ 	\int\limits_0^{2\pi} {d \varphi_1\over 2\pi} P(\varphi_1,\bm q) \, 
\left[ W^{ss}_{aa}(\theta_1) \mp W^{sa}_{as}(\theta_1) \right] C_\pm(\varphi_1,\varphi_{\bm k'},\bm q),
\nonumber
\end{eqnarray}

\begin{eqnarray}
\label{C1}
&&	C_1(\varphi_{\bm k},\varphi_{\bm k'},\bm q) = W^{ss}_{ss}(\theta) \\
	&+&  	\int\limits_0^{2\pi} {d \varphi_1\over 2\pi} P(\varphi_1,\bm q) \, 
\biggl[ W^{ss}_{ss}(\theta_1) C_1(\varphi_1,\varphi_{\bm k'},\bm q) 
\nonumber \\
	&&
+ W^{sa}_{sa}(\varphi_{\bm k},\varphi_1) C_2(\varphi_1,\varphi_{\bm k'},\bm q)\biggr] ,
\nonumber
\end{eqnarray}

\begin{eqnarray}
\label{C2}
&&	C_2(\varphi_{\bm k},\varphi_{\bm k'},\bm q) = W^{as}_{as}(\theta) \\
	&+&  	\int\limits_0^{2\pi} {d \varphi_1\over 2\pi} P(\varphi_1,\bm q) \, 
\biggl[ W^{as}_{as}(\varphi_{\bm k},\varphi_1) C_1(\varphi_1,\varphi_{\bm k'},\bm q) 
\nonumber \\
	&&
+ W^{aa}_{aa}(\theta_1) C_2(\varphi_1,\varphi_{\bm k'},\bm q)\biggr].
\nonumber
\end{eqnarray}
Here we introduced $\theta=\varphi_{\bm k}-\varphi_{\bm k'}$ and  $\theta_1=\varphi_{\bm k}-\varphi_1$.

We have found the Cooperons $C_+$, $C_-$, $C_1$ and $C_2$ expanding them in Fourier series which transfers the integral Eqs.~\eqref{C_pm}-\eqref{C2} to systems of linear equations. Since the correlators have a finite number of Fourier harmonics, the linear equation systems are finite: 
7 independent equations for both $C_+$ and $C_-$, and 14 coupled equations for $C_1$ and $C_2$.
The Cooperon Fourier harmonics have been found by a numerical solution of these linear equation systems.

The found Cooperons allowed us to calculate the WL 
conductivity correction $\sigma$. It is well known that this correction is given by a sum of two maximally crossed diagrams. Accordingly, it equals to a sum of two terms, $\sigma=\sigma_a+\sigma_b$, where $\sigma_a$ arises from the backscattering processes and $\sigma_b$ contribution is due to coherent scattering by arbitrary angles. These corrections are expressed via the Cooperons as follows
\begin{eqnarray}
	\sigma_a &=& - {e^2(v_{\rm F} \tau_1)^2\over 2\pi\hbar} \int {d^2q\over (2\pi)^2} \int\limits_0^{2\pi} {d \varphi\over 2\pi} P(\varphi,\bm q)
\\
&\times& \biggl\{ C_-^{(3)}(\varphi,\varphi+\pi,\bm q) - C_+^{(3)}(\varphi,\varphi+\pi,\bm q) \nonumber
\\ 
 && + 2 {\rm Re} \left[ C_1^{(3)}(\varphi,\varphi+\pi,\bm q)\right] 
 \biggr\},
 \nonumber
\end{eqnarray}
\begin{eqnarray}
&& \sigma_b = 
 - {e^2(v_{\rm F} \tau_1)^2\over \pi\hbar} 
 \\ 
&\times& \int {d^2q\over (2\pi)^2} \int\limits_0^{2\pi} {d \varphi\over 2\pi} P(\varphi,\bm q) \cos{\varphi} \int\limits_0^{2\pi} {d \varphi'\over 2\pi} P(\varphi',\bm q) \cos{\varphi'} 
 \nonumber \\
&\times& \biggl\{ 
 \sum_\pm \left[ W^{ss}_{aa}(\theta) \mp W^{sa}_{as}(\theta) \right] C_\pm^{(2)}(\varphi,\varphi',\bm q) 
 \nonumber \\
 && + 2{\rm Re} \left[ W^{ss}_{ss}(\theta) C_1^{(2)}(\varphi,\varphi',\bm q) 
\right]
 \nonumber \\
 && + 2{\rm Re} \left[ W^{sa}_{sa}(\varphi,\varphi')C_2^{(2)}(\varphi,\varphi',\bm q) \right]
 \biggr\}.
  \nonumber
  \end{eqnarray}
Here $\theta=\varphi-\varphi'$, the upper index (2) and (3) indicates that the diagrams are started from 2 and 3 impurity lines, and the factor $\tau_1^2$ comes from the renormalization of two vertexes due to the scattering anisotropy.

\begin{figure}[t!]
\includegraphics[width=0.95\linewidth]{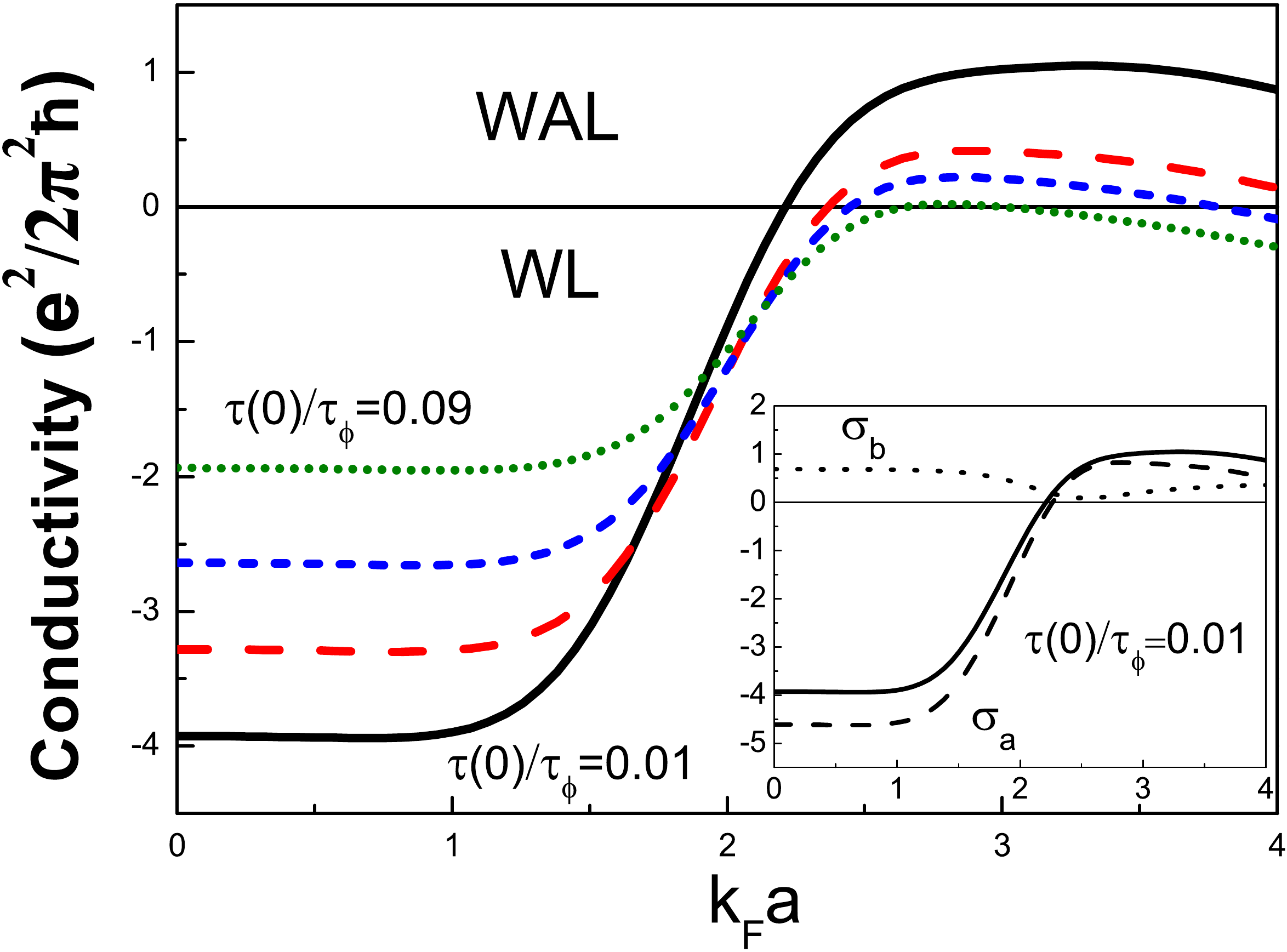} 
\caption{Conductivity correction as a function of hole density at 
$\tau(0)/\tau_\phi=0.01, 0.02, 0.04$ and 0.09. 
Inset: the backscattering ($\sigma_a$) and non-backscattering ($\sigma_b$) contributions.}
\label{fig_sigma}
\end{figure}

The results of calculation of the WL conductivity correction is shown in Fig.~\ref{fig_sigma}. The WL to WAL transition with increase of the hole density is clearly seen for $\tau(0)/\tau_\phi \leq 0.04$. It occurs at $k_{\rm F}a \approx 2$ which corresponds to the density $p \approx 7 \times 10^{11}$~cm$^{-2}$ for the quantum well width $a=100$~\AA. However, Fig.~\ref{fig_sigma} demonstrates that the conductivity is a non-monotonous function of the hole density which changes sign and become again negative at high $k_{\rm F}a$ for shorter dephasing times $\tau_\phi$.
Inset in Fig.~\ref{fig_sigma} shows the backscattering and nonbackscattering contributions, $\sigma_a$ and $\sigma_b$.

The backscattering contribution dominates for all values of the parameter $k_{\rm F}a$. This can be  explained by the approximate relation 
\[
\sigma_b \approx -{\tau_1-\tau_0\over\tau_1}\sigma_a,
\] 
which is derived in the diffusion approximation.~\cite{JETP98,FTP98}
The difference between $\tau_1$ and $\tau_0$ is nonzero, but it does not exceed 30\%, see Fig.~\ref{fig_disp_times}. Therefore the nonbackscattering contribution $\sigma_b$ does not reach the large value $(e^2/2\pi^2\hbar)\ln{(\tau_\phi/\tau_1)}$. In contrast, it does not exceed the value at $k=0$ where $\sigma_b = (e^2/2\pi^2\hbar)\ln{2}$, see Fig.~\ref{fig_sigma}.

\begin{figure}[t!]
\includegraphics[width=0.9\linewidth]{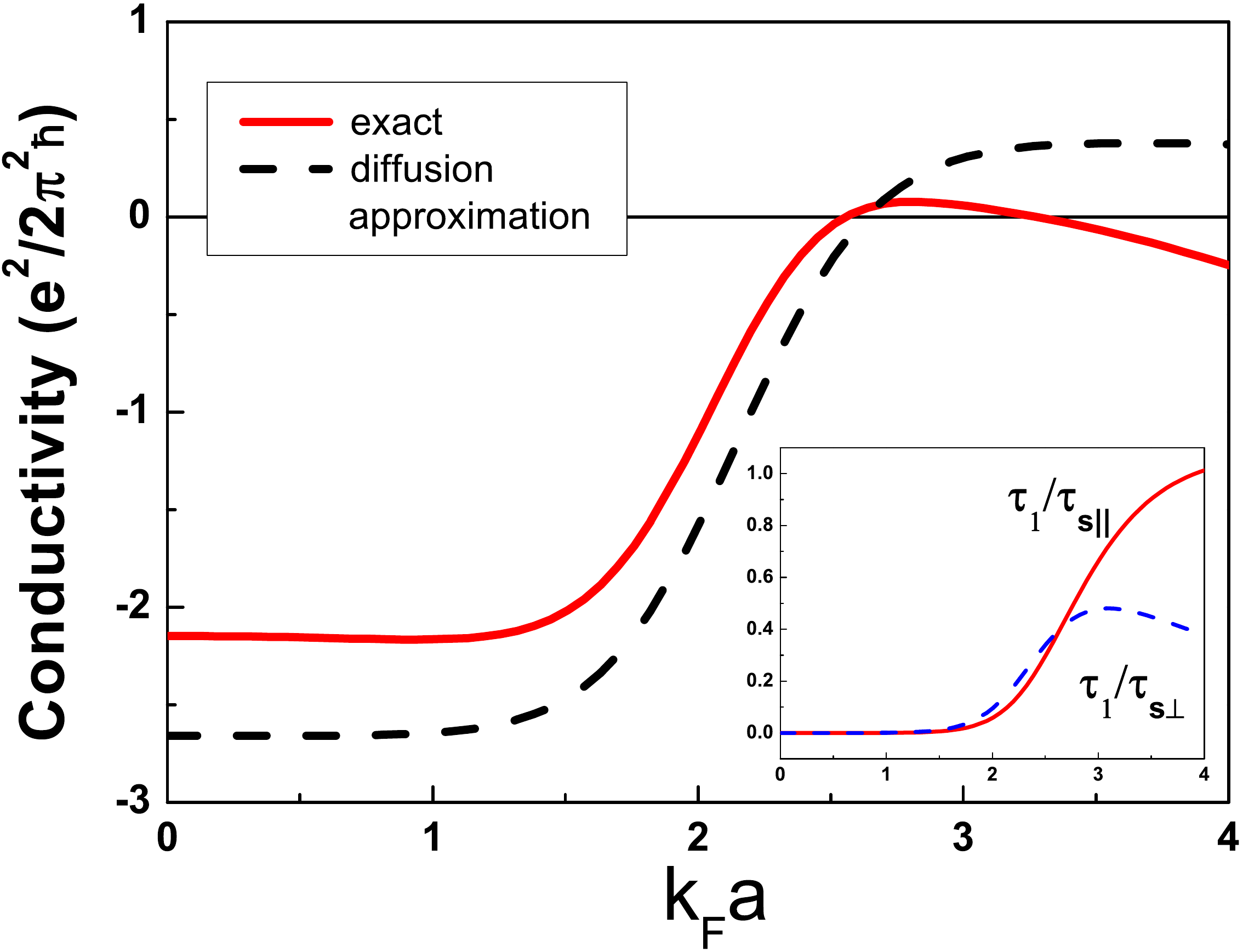} 
\caption{The backscattering correction at $\tau(0)/\tau_\phi=0.07$ calculated exactly (solid line) and in the diffusion approximation (dashed line). Inset: spin relaxation rates.}
\label{fig_compar_diff}
\end{figure}

In order to explain qualitatively the non-monotonous behaviour of the conductivity correction with the hole density we also compare the results of our exact calculation with ones obtained in the diffusion approximation:
%
%The behaviour of $\sigma_a$ can  be explained qualitatively by the expression obtained in the diffusion approximation:
\begin{equation}
\label{diff}
	\sigma_a \approx  {e^2\over 4\pi^2 \hbar} \left[ 2\ln{\left({\tau_1\over\tau_{s\parallel}}+{\tau_1\over\tau_\phi}\right)} + \ln{\left({\tau_1\over\tau_{s\perp}}+{\tau_1\over\tau_\phi}\right)} - \ln{\tau_1\over\tau_\phi}\right].
\end{equation}
Here the longitudinal and the transverse spin relaxation rates are given by:~\cite{SOI_MIT}
\[
{1\over \tau_{s\parallel}} = {1\over \tau_0} - {2\pi \over \hbar} g \int\limits_0^{2\pi} {d\theta \over 2\pi} \left< V_{ss}^2(\theta) \right>,
\]
\[
{1\over \tau_{s\perp}} = {4\pi \over \hbar} g \int\limits_0^{2\pi} {d\theta \over 2\pi} \left< |V_{sa}(\theta)|^2 \right>.
\]
The dependence Eq.~\eqref{diff} is shown in Fig.~\ref{fig_compar_diff} together with the result of the exact calculation. The increase of the WL correction at $0<k_{\rm F}a<3$ is described by fast increase of both spin relaxation rates which makes smaller two first logarithmic terms in Eq.~\eqref{diff}. This demonstrates that WL to WAL transition in hole systems at long dephasing times is driven by increase of spin-orbit coupling with increase of the hole density. However, at $k_{\rm F}a>3$ the WL correction calculated in the diffusion approximation saturates at the value smaller than $(e^2/4\pi^2\hbar)\ln{(\tau_\phi/\tau_1)}$ obtained in the limit of very fast spin relaxation. 
This is caused by decrease of the transverse spin relaxation rate $1/\tau_{s\perp}$ and by saturation of the longitudinal rate $1/\tau_{s\parallel}$, 
see inset to Fig.~\ref{fig_compar_diff}.

This behaviour of the spin relaxation rates at $ka>1$ can be explained by considering limit of a very wide quantum well. 
The coefficients $v_{0\ldots 3}$ in Eqs.~\eqref{s},~\eqref{a}  in this limit are $v_0 = 1/2$, $v_1 = v_3 = 0$, $v_2 = \sqrt{3}/2$,~\cite{MPP_91} so the symmetrical and asymmetrical wave functions have the following forms:
\begin{eqnarray}
	F_{s,a} &=& \cos{\left( {\pi z\over a} \right)} { \exp{(\pm {\rm i}\varphi_{\bm k})} \over \sqrt{a}} \nonumber \\ 
	&\times& \left[ (Y\cos{\varphi_{\bm k}} - X\sin{\varphi_{\bm k}}) s_\pm
	+ {\rm i}Z {\rm e}^{\pm {\rm i}\varphi_{\bm k}} s_\mp \right]. \nonumber
\end{eqnarray}
These wave functions correspond to the angular momentum projections $\pm 3/2$ on the direction of $\bm k$ in the structure plane as it should be at large $ka$. 
Calculating the scattering matrix elements $V_{ss} = V_0\langle F_{s\bm k'}|F_{s\bm k}\rangle$ and $V_{sa} = V_0\langle F_{s\bm k'}|F_{a\bm k}\rangle$ we derive that at $ka \to \infty$
\[
1/\tau_{s\perp} = 0, \qquad
0.9 \tau_{s\parallel} = \tau_0=\tau_1.
\]
Infinite $\tau_{s\perp}$ shows that the hole spin relaxation is suppressed not only at small but also at large $ka$. According to the calculation results shown in inset to Fig.~\ref{fig_compar_diff} this limiting case is almost realized at $k_{\rm F}a\leq 4$ when only the ground subband is occupied.
Therefore the first logarithmic term in Eq.~\eqref{diff} is small at $k_{\rm F}a\sim 4$ while the second one increases  partially compensating the last term, and the total correction saturates.
Moreover, in contrast to the diffusion-approximation result, the exactly calculated WL correction totally repeats the behaviour of $\tau_1/\tau_{s\perp}$:  the above derived suppression of hole spin relaxation in wide quantum wells is reproduced in the WL conductivity correction as it is seen from exact calculation result shown Fig.~\ref{fig_compar_diff} which has a maximum and decreases at larger $ka$.

\begin{figure}[t!]
\includegraphics[width=0.9\linewidth]{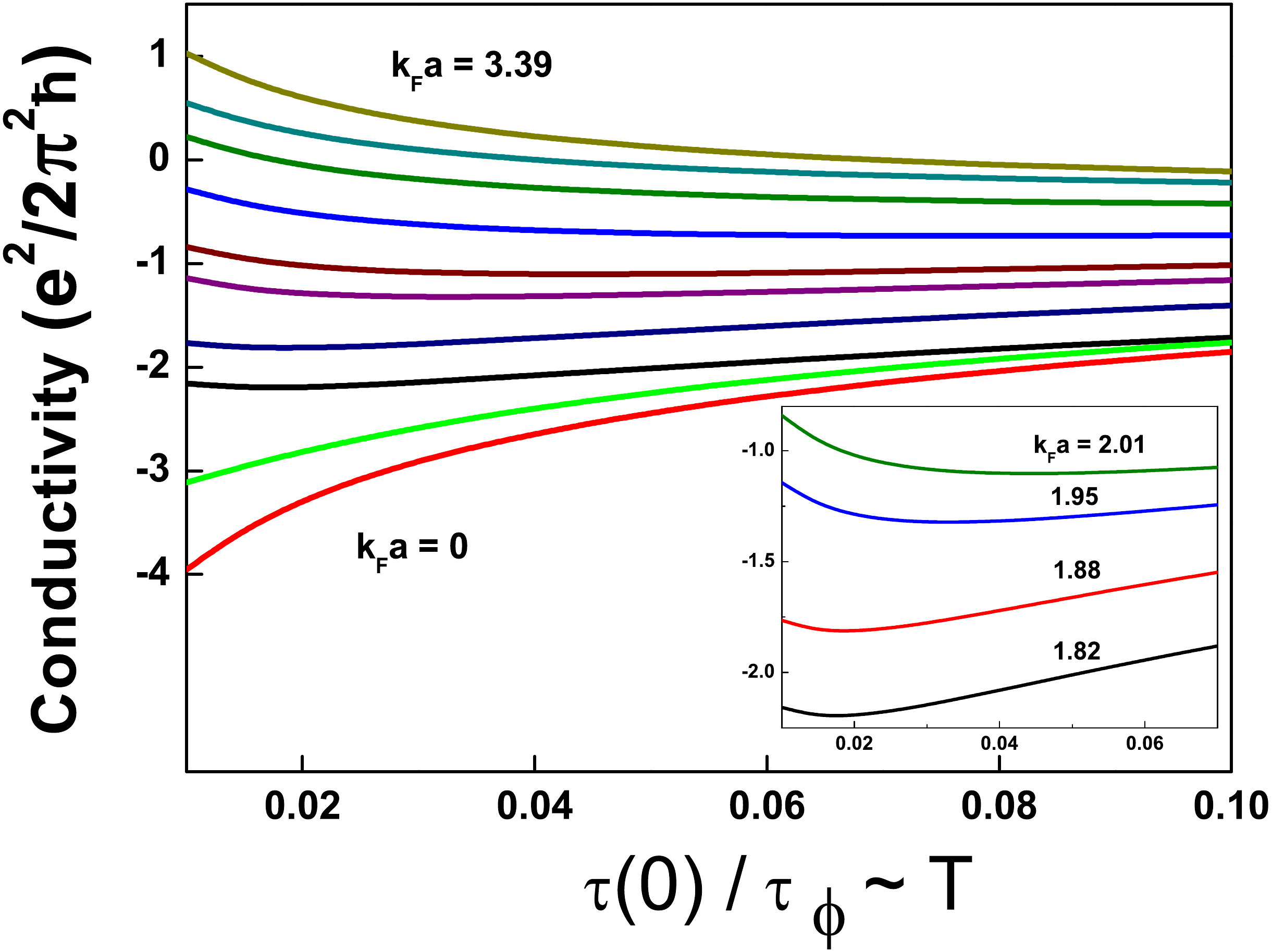} 
\caption{The temperature dependence of the conductivity correction at different hole densities. Inset highlights the region where the conductivity has a minimum.}
\label{fig_temp_dep}
\end{figure}

We have also calculated the WL conductivity correction as a function of temperature $T \sim \tau_\phi^{-1}$ at different hole densities. The result shown in Fig.~\ref{fig_temp_dep} demonstrates the transition from metal to insulator temperature behaviour. This transition is due to spin-orbit interaction which is present even in symmetrical quantum wells with the degenerate hole energy spectrum. Inset to Fig.~\ref{fig_temp_dep} shows non-monotonous temperature dependence at moderate hole densities.

To summarize, we have developed WL theory for holes in quantum wells taking into account 
both ballistic and diffusion processes. 
%Contributions to the conductvity correction due to both backscattering and nonbackscattering processes are taken into account. 
The conductivity correction is calculated in a wide range of hole densities and temperatures taking into account real nonparabolicity of the energy spectrum. 
WL instead of WAL can occur at high densities due to suppression of hole spin relaxation at large in-plane wavevectors.
The transition from metal to insulating temperature behaviour is demonstrated in the symmetrical quantum well with spin-degenerate hole subbands. Such transition can be detected by magnetoconductivity measurements. This mechanism can concur with the mechanism due to Rashba splitting of hole energy spectrum which has been used in Ref.~\onlinecite{Minkov_strained} for explanation of WL experiments  on two-dimensional hole systems.

{\it Acknowledgments.} 
We thank M.~M. Glazov, M.~O.~Nestoklon and E.~L.~Ivchenko for stimulating discussions. The work was supported by RFBR, President grant for young scientists, Program ``Leading Scientific Schools'' (\#5442.2012.2), and ``Dynasty'' Foundation -- ICFPM.

\end{document}